# Multiple Mobile Target Detection and Tracking in Active Sonar Array Using a Track-Before-Detect Approach

Avi Abu$^§$, Nikola Mišković$^♯$, Oleg Chebotar$^§$, Neven Cukrov$^H$, Roee Diamant$^{§,♯,C,\star}$

$^§$Hatter Department of Marine Technologies, University of Haifa, Israel

$^♯$Faculty of Electrical Engineering and Computing, University of Zagreb, Croatia

$^H$Division for marine and environmental research, Ruder Bošković Institute, Croatia

$^C$Project CETI, New York, NY 10003, USA

$^\star$Corresponding author, email: `roee.d@univ.haifa.ac.il`

*Abstract*—We present an algorithm for detecting and tracking underwater mobile objects using active acoustic transmission of broadband chirp signals whose reflections are received by a hydrophone array. The method overcomes the problem of high false alarm rate by applying a track-before-detect approach to the sequence of received reflections. A 2D time-space matrix is created for the reverberations received from each transmitted probe signal by performing delay and sum beamforming and pulse compression. The result is filtered by a 2D constant false alarm rate (CFAR) detector to identify reflection patterns corresponding to potential targets. Closely spaced signals for multiple probe transmissions are combined into blobs to avoid multiple detections of a single object. A track-before-detect method using a Nearly Constant Velocity (NCV) model is employed to track multiple objects. The position and velocity is estimated by the debiased converted measurement Kalman filter. Results are analyzed for simulated scenarios and for experiments at sea, where GPS tagged gilt-head seabream fish were tracked. Compared to two benchmark schemes, the results show a favorable track continuity and accuracy that is robust to the choice of detection threshold.

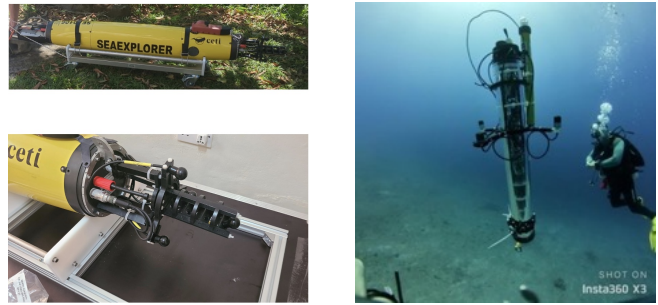

(a) The CETI glider (built by Alseamar) with array of hydrophones embedded in the nose section for deep-water squid detection. Active acoustics performed by an acoustic pinger strapped to the glider. Analysis is performed onboard by a Jetson board, and online change of mission is possible by a backseat driver.

(b) The SOUND floater with array of hydrophone on its side and a projector on top. Analysis is performed on an STM board within the floater, and detections are shared throughout a swarm of floaters for improved SNR and response.

Fig. 1. Platforms used for the task of tracking mobile submerged targets.

## I. INTRODUCTION

The detection and tracking of underwater objects is required for applications such as the monitoring of marine fauna [1], the identification of schools of fish in commercial ventures [2] and the detection of prey of sperm whales (*Physeter macrocephalus*) for behavioral analysis [3]. Monitoring is done from vessels and underwater by floaters or gliders. Real-time analysis enables an immediate response to either report a detection event or approach the source to achieve a better signal-to-noise ratio (SNR). Two specific applications we are focusing on are the monitoring of deep-sea giant squid (*Architeuthis dux*) by autonomous gliders to draw conclusions about the behavior of sperm whales, for which squid are a major food source, and the monitoring of fish populations for biodiversity analysis. This is needed both for regulators who need to set data-driven policy and for sustainable fisheries to reduce bycatch by making fishermen aware of the location of fish schools. The first application is considered within a project called *CETI*, where we have set the ambitious goal of understanding the language of sperm whales by tracking their activity and vocalizations. The second application is carried out as part of a project called *SOUND*, where we deploy a swarm of autonomous floaters to monitor the water column in search of schools of fish. In both cases, the monitoring is done by active acoustics. In particular, by emitting acoustic signals and analyze their reflections to find moving targets. Pictures of our glider and floater platforms can be found in Fig 1a and 1b respectively. In this paper, we focus on the task of target detection and localization.

The detection of underwater objects is dominated by the signal-to-clutter ratio (SCR), which includes all reflections that

This research was supported by the Schmidt Marine Foundation via the Global Fisheries Tech Initiative, by the Horizon Europe programme of the European Union under the UWIN-LABUST project (project number 101086340), and by Project CETI via grants from Dalio Philanthropies and Ocean X; Sea Grape Foundation; Rosamund Zander/Hansjorg Wyss, Chris Anderson/Jacqueline Novogratz through The Audacious Project: a collaborative funding initiative housed at TED.

do not associate with the target. Due to the low target strength of fish (in the order of -40 dB, [4]), the SCR is likely to be in the order of 0 dB. Therefore, in active acoustic detection, an array of hydrophone receivers is used to utilize the spatial domain by e.g., beamforming [5]. In our case, we use an array of four hydrophones positioned in the plane. Here, the challenge is to handle the ambiguity in the angle estimation, which reaches 15° for a plenary array [6]. Another challenge lies in the fast mobility of the tracked object and the possible existence of multiple targets.

Our multi-target detection and localization solution is based on the track-before-detect approach to analyze the reflections of a sequence of broadband chirp signals received by an acoustic array. For each transmitted chirp signal, a 2D angle-distance matrix is formed by pulse compression and delay-and-sum beamforming. The samples of this matrix are thresholded by a constant false alarm detector (CFAR) to eliminate weak reflections that may not be associated with the search targets. The remaining samples are grouped into blobs, followed by further filtering to remove detected regions that are too small or too large. To track objects, we employ the debiased converted measurement Kalman filter [7] with a near constant velocity dynamical model(NCV). The result is a scheme for 2D target tracking whose main advantage is the ease of handling objects with low SCR in a high false alarm rate environment.

The main contributions of this paper are twofold:
1) A 2D multitarget scheme for identifying moving fish by active acoustics while attenuating noise.
2) Merging adaptive beamforming and tracking into a single track-before-detect approach to reduce the false alarm rate. Incorporating the beamformer into a track-before-detect scheme makes it possible to overcome low SCR values while utilizing the expected smooth motion of the target.

The results of numerical simulations show the applicability of our method to overcome low SCR values in multi-target scenarios and its advantages over two benchmark schemes. The results of experiments at sea, where live fish were tagged with a GPS receiver and detected by our float, show the applicability of our approach in realistic scenarios and the accuracy achieved in a real marine environment.

The remaining of this paper is organized as follows. In Section II, we survey the state-of-the-art for object detection and tracking of underwater objects. System model and main assumptions are presented in Section III. Our target detection scheme is introduced in Section IV, and our multitarget tracking is outlined in Section V. Numerical and experimental results are discussed in Section VI. Concluding remarks are offered in Section VII.

## II. Literature Review

Target detection using the constant false alarm rate (CFAR) method is a common application for underwater monitoring tasks [8]. For clutter mitigation, a cell-averaged CFAR detector is offered in [9], where multiple cells are averaged to set the detection threshold. Other similar CFAR approaches are the accumulated cell-averaged CFAR [10], in which the accumulated sample matrix is calculated to reduce the computational cost in calculating the detection threshold, and the order statistics CFAR [11], in which noisy samples are attenuated by sorting the samples and selecting only the dominant samples. A detection method that is robust to complex environments is provided in [12], where the 2D matrix is clustered using the K-Means algorithm and the objects are identified using the resulting binary matrix.

For target tracking, the extended Kalman filter [13] is used due to its simplicity in tracking nonlinear dynamic or measurement models. The method relies on a motion model for the target, and a linearization of the measurement model. Versions of Kalman tracking for object tracking is the debiased converted measurement Kalman filter [7], in which the non-linear polar measurements are converted into a linear system of Cartesian coordinates taking into account the bias due to the nonlinearity of the conversion.

For multitarget tracking, data association of detections into tracks is necessary. Methods include the multi-hypothesis tracking (MHT), in which all possible mappings between detections and tracks are considered, and joint probabilistic data association (JPDA) [14], in which a track is assigned to multiple detections by a probability measure. To reduce the complexity of the MHT filter, a Gaussian mixture probability hypothesis density (GM-PHD) filter is introduced in [15]. This filter has closed-form recursions for the propagation of the state vector and its covariance and its advantage lies in its practicability for underwater scenarios with multiple targets. In [16], CA-CFAR with threshold segmentation is used to reduce false alarms. The method applies iterative threshold segmentation to reduce false alarms. Another multi-target approach for underwater applications is the sequential Monte Carlo density algorithm [17]. For object detection, the authors offered a combination of CA-CFAR and the K-means detector. To reduce the number of missed tracks, a continuous lost frame threshold is added. More specifically, if the predicted states of a track are not correlated with an object across multiple sonar frames, the state vector and its covariance are replaced by the last updated state of the track. However, the results are limited to high SCR.

Acoustic arrays are used to increase the SCR. In [18], detection and tracking of underwater objects is performed using a multi-beam sonar. Detection is performed by segmenting the sonar image, while the position of the object after detection is determined by its contour position. Object tracking is done using a particle filter to model the posterior by a series of weighted samples. In [19], the Gaussian mixture cardinalized probability hypothesis density filter is applied to multistatic sonar data. The posterior is estimated by a Gaussian mixture and the state vector and its covariance are predicted by the EKF. The Gaussian mixture is also used in [20] for a cardinalized likelihood hypothesis filter over real sonar data obtained from a sea trial. This filter has the advantage that it can well handle clutter. A linear motion model is used with a non-linear measurement model. However, due to the high false alarm rate in real sonar data, there is a large track breakage rate. In [21], three optimization algorithms for probabilistic maximum likelihood data association are introduced. The formalization includes the multi-pass grid search, the genetic algorithm and the developed directed subspace search. However, solving directly involves high computational costs. Although the aforementioned methods provide good results, the complexity in tracking or detection is high and the implementation on an on-board computer may be limited to short missions. We identify two remaining challenges for the task of multitarget detection and tracking multiple targets: 1) high complexity and 2) high rate of missing tracks. The latter is especially true in scenarios with high false alarm rates and low SCR environment.

## III. System Model

Our goal is to detect and track mobile underwater objects while achieving a favorable trade-off between precision and recall as well as high accuracy in target localization. To manage a small array suitable for small platforms such as a floater or a glider, we focus only on 2D localization. Our scenario involves a Lagrangian floater or glider that drifts with the water current, changing only its depth to a target profile. This can be achieved by a piston-based buoyancy control, as is the case with the system used in our results below, or by a bladder [22].

For active acoustic detection, we use a single projector that emits a sequence of chirp signals of large bandwidth-duration product for high pulse compression. The reflections are recorded by an array of $N$ hydrophones ($N = 4$ in our results) mounted in a plenary array around the platform, as shown in Fig. 1. Further details can be found in [23]. The data received from the $N$ sensors is transferred to the angle-distance domain by applying a matched filter followed by a beamforming. The result is a 2D matrix in which each sample is associated with a time-of-arrival, which can be converted to a distance if the speed of sound is known, and an angle of arrival. The time domain input of the $m-th$ hydrophone with $N$ sources is modeled as follows

$$s_m(t) = \sum_{n=1}^{N} d(t - \tau_m(\theta_n)) + \eta_m(t) , \quad (1)$$

where $d(t)$, $\tau_m(\theta_n)$, and $\eta_m(t)$ stand for the signal waveform, the time delay of the $n-th$ source with azimuth denoted by $\theta_n$, and the clutter noise. The time delay is given by

$$\tau_m(\theta_n) = \frac{1}{c} \bar{p}_m \cdot \bar{\omega}(\theta_n) , \quad (2)$$

where $c$ is sound velocity at water, $\bar{p}_m$ is the Cartesian location of the $m-th$ hydrophone, and $\bar{\omega}(\theta_n)$ is the spatial direction of the $n-th$ source given by

$$\underline{\omega}(\theta_n) = \begin{bmatrix} \cos(\theta_n) \\ \sin(\theta_n) \end{bmatrix} . \quad (3)$$

We assume that the speed of sound in water is known or measured by the profiling platform. Following [24], we assume that the angle-distance samples are independent and identically distributed (i.i.d.) and follow the Rayleigh distribution. We assume that the polar measurements are uncorrelated and corrupted by an additive white Gaussian noise with known or measured covariance. We neglect multipath reflections from the target through the seafloor and assume that the dominant reflection is given by the direct path. We argue that this is a reasonable assumption since the planar array does not account for elevation angle and most reflections are received from the same bearing. The dynamic path of the targets is assumed to follow a model with nearly constant velocity. The process noise of the dynamic target model is assumed to be Gaussian noise with zero mean and a known standard deviation $\sigma_\zeta$.

## IV. Underwater Target Detection

Referring to the block diagram in Fig. 2, our target detection involves four steps: 1) pulse compression using a matched filter, 2) beamforming $U$ spatial domain beams and forming an angle-distance 2D matrix, 3) applying a constant 2D false alarm rate (CFAR) detector, and 4) binary map extraction and blob detection.

### A. Forming the 2D angle-distance Matrix

For each emission of chirp signals, we create a 2D angle-distance matrix from the received reflections. The process of creating these matrices is as follows. First, a buffer of length $L_s$ is extracted at each of the array elements. Multiplying $L_s$ by half the speed of sound reflects the maximum detection range of the system. The received buffer is filtered separately at the output of each hydrophone with a matched filter $f(t)$ that is adapted to the transmitted chirp signal. The compressed signals are beamformed into $u = 1, \ldots, U$ beams using the delay-and-sum procedure [25]. Formally, for a beam azimuth $\phi_u$, the associated delay is $\tau_m(\phi_u)$. The data after beamforming for a reflection received at time instance $t_s$, $s = 1, \ldots, L_s$, $y_u(t_s)$, is given by

$$y_u(t_s) = \sum_{m=1}^{M} s_m(t_s + \tau_m(\phi_u)) . \quad (4)$$

The angle-distance matrix $I \subseteq \mathbb{R}^{U \times L_s}$ is obtained by

$$I = \begin{pmatrix} y_1(t_1) & \cdots & y_1(t_{L_s}) \\ \vdots & \cdots & \vdots \\ y_U(t_1) & \cdots & y_U(t_{L_s}) \end{pmatrix} . \quad (5)$$

### B. Blob Detection

The corresponding angle-distance 2D matrix $I$ for each emission is filtered to reduce clutter. The detection threshold is calculated according to a 2D CFAR detector to yield a binary matrix samples that passed ('1') or didn't pass ('0') the threshold. The process involves estimation of the 2D angle-distance sample distribution parameters to calculate the CFAR detection threshold. The Rayleigh distribution,

$$\mathcal{R}_\alpha(I(u,v)) = \frac{I(u,v)}{\alpha^2} \exp\left(-\frac{I^2(u,v)}{2\alpha^2}\right), \ I(u,v) > 0 , \quad (6)$$

where $\alpha$ is the law-specific parameter, is adopted to model the clutter distribution [24]. Parameter $\alpha$ is estimated by the maximum-likelihood [26]

$$\alpha = \sqrt{\frac{1}{2|\Omega|} \sum_{\{u,v\} \in \Omega} I^2(u,v)} , \quad (7)$$

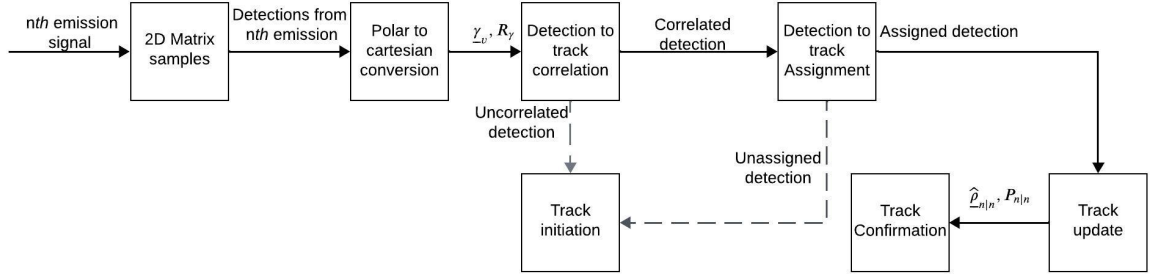

Fig. 2. A block diagram of the proposed detection and tracking scheme.

where $\Omega$ stands for the group of samples surrounding the sample under test in $I$ that belong to the reference region. The CFAR threshold $T_c$ is given by

$$T_c = \sqrt{\left(P_{fa}^{-\frac{1}{|\Omega|}} - 1\right) \sum_{\{u,v\} \in \Omega} I^2(u, v)} , \quad (8)$$

where $P$fa is a target probability of false alarm.

The CFAR detection binary map $B \in [0, 1]$ is obtained by

$$B(u, v) = \begin{cases} 1, & \text{if } I(u, v) > T_c \\ 0, & \text{else} \end{cases} . \quad (9)$$

Adopting the 4-connectivity method [27], we identify in $B$ several regions of interest (ROI) who contain potential targets and that form a connected region. We refer to such region as a *blob*. The range $r_l$ and azimuth $\theta_l$ of the $l$-th blob is estimated by

$$r_l = \frac{\bar{v}_l}{f_s} c , \quad (10)$$

where $\bar{v}_l$ is the $v$-th coordinate of the center mass of the $l$-th blob, and $f_s$ is the sampling frequency. The azimuth $\theta_l$ is given by

$$\theta_l = \arg\max_u I(u, \bar{v}_l) . \quad (11)$$

When the number of elements in the acoustic array is small, the beamwidth is large. As a result, the beamforming grating lobes are also large, and a target can be received at adjacent beams. To manage this ambiguity, we merge blobs that are closely spaced. Let $\mathcal{L}$ be the group of all blobs obtained from a single emission. Let $l(i), l(j) \in \mathcal{L}$, be two separate blobs, which are merged if

$$|r_{l(i)} - r_{l(j)}| < \psi_r, \quad (12a)$$
$$|\theta_{l(i)} - \theta_{l(j)}| < \psi_\theta, , \quad (12b)$$

where $\psi_r$ and $\psi_\theta$ are the merging thresholds in range and azimuth, respectively. The threshold in range can be obtained by utilizing knowledge about the expected size of the mobile target to be detected, and the threshold in azimuth can be set by the position of the grating lobes of the array. The range and azimuth of the $i$-th merged blob at the emission number $n$ is given by

$$r_n^i = \frac{1}{|\mathcal{E}|} \sum_{l \in \mathcal{E}} r_l \quad (13a)$$

$$\theta_n^i = \frac{1}{|\mathcal{E}|} \sum_{l \in \mathcal{E}} \theta_l, \quad (13b)$$

where $\mathcal{E}$ is the group of all blobs that fulfill conditions in (12). A tracking process is now applied on the merged blobs to detect a mobile target with a nearly constant velocity dynamical model. To keep the tracking process efficient with low complexity, we apply a linear Kalman filter, with a debiased polar to Cartesian measurement conversion as a fast and efficient method to handle a non linear measurement model.

### C. Polar to Cartesian Conversion

Following [7], we convert the polar measurements $(r_n^i, \theta_n^i)$ into a Cartesian space. The covariance matrix of the polar measurement noise is

$$R = \begin{bmatrix} \sigma_r^2 & 0 \\ 0 & \sigma_\theta^2 \end{bmatrix} , \quad (14)$$

where $\sigma_r$ and $\sigma_\theta$ are the standard deviations of the range and the azimuth measurements, respectively. The debiased converted measurement $\underline{\gamma}_n^i$ in Cartesian coordinates is given by [7]

$$\underline{\gamma}_n^i = \begin{bmatrix} r_n^i \cos(\theta_n^i)\left(1 - e^{-\sigma_\theta^2} - e^{-0.5\sigma_\theta^2}\right) \\ r_n^i \sin(\theta_n^i)\left(1 - e^{-\sigma_\theta^2} - e^{-0.5\sigma_\theta^2}\right) \end{bmatrix} . \quad (15)$$

## V. UNDERWATER TARGET TRACKING

The final step in our target detection is tracking, where we confirm valid mobile targets through detected blobs that follow a dynamical model of constant velocity. The dynamical model is given by

$$\bar{\rho}_n = A\bar{\rho}_{n-1} + G\bar{\zeta}_{n-1} , \quad (16)$$

where $\bar{\rho}_n = [\rho_x \ \rho_y \ \dot{\rho}_x \ \dot{\rho}_y]^T$, $\rho_x$ and $\rho_y$ are the position vector of a target in X and Y cartesian frame and $\dot{\rho}_x$, $\dot{\rho}_y$ are the Cartesian velocities, respectively, and $\bar{\zeta}_{n-1}$ is the process noise. Matrix $A$ in (16) is given by

$$A = \begin{bmatrix} 1 & 0 & t_n - t_{n-1} & 0 \\ 0 & 1 & 0 & t_n - t_{n-1} \\ 0 & 0 & 1 & 0 \\ 0 & 0 & 0 & 1 \end{bmatrix}, \quad (17)$$

matrix $G$ is given by

$$G = \begin{bmatrix} \frac{(t_n - t_{n-1})^2}{2} & 0 \\ 0 & \frac{(t_n - t_{n-1})^2}{2} \\ t_n - t_{n-1} & 0 \\ 0 & t_n - t_{n-1} \end{bmatrix}, \quad (18)$$

and $t_n$ is the time of the $n$-th blob.

The measurement equation is given by

$$\bar{\gamma}_n^i = H\bar{\rho}_n + \bar{\kappa}_n, \quad (19)$$

with

$$H = \begin{bmatrix} 1 & 0 & 0 & 0 \\ 0 & 1 & 0 & 0 \end{bmatrix}, \quad (20)$$

and $\bar{\kappa}_n$ is the measurement noise. The goal of our tracking is to estimate the position and velocity trajectories of a target. For this, we use Kalman filtering. The Kalman filter includes predict and update steps. The predicted step of the state vector and its covariance for the $k$-th track are given by

$$\widehat{\bar{\rho}}_{n|n-1}^k = A\widehat{\bar{\rho}}_{n-1|n-1}^k \quad (21a)$$

$$P_{n|n-1}^k = AP_{n-1|n-1}^k A^T + Q, \quad (21b)$$

where $Q = \sigma_\zeta^2 GG^T$ is the process noise covariance matrix.

The update stage of the state vector and its covariance matrix is given by:

$$\widehat{\bar{\rho}}_{n|n}^k = \widehat{\bar{\rho}}_{n|n-1}^k + K_n\left(\bar{\gamma}_n^i - H\widehat{\bar{\rho}}_{n|n-1}^k\right), \quad (22a)$$

$$P_{n|n}^k = P_{n|n-1}^k + K_n\left(W_n\right)^{-1} K_n^T, \quad (22b)$$

, where

$$K_n = P_{n|n-1}^k H^T \left(HP_{n|n-1}^k H^T + R_\gamma\right)^{-1}, \quad (23)$$

and $W_n = \left(HP_{n|n-1}^k H^T + R_\gamma\right)$. The term $R_\gamma$ is the covariance of the debiased converted measurement noise [24].

### A. Detection to Track Correlation

To decrease computation time, we first apply a coarse gating between tracks and blobs that originate from one emission. Formally, only a blob that fulfills the following two conditions with at least one track is allowed, otherwise a new track is created. The gating between the $i$-th blob and the $k$-th track in range and azimuth is given by

$$\left|\widehat{r}_{n-1}^k - r_n^i\right| < G_r \quad (24a)$$

$$\operatorname{atan2}\left(\widehat{\rho}_{x,n-1}^k, \widehat{\rho}_{y,n-1}^k\right) - \theta_n^i < G_\theta, \quad (24b)$$

where $G_r$ and $G_\theta$ are the gating thresholds in range and azimuth, respectively. These thresholds can be set by utilizing the knowledge about the maximum velocity of the mobile target and the time between adjacent emissions. Terms $\widehat{r}_{n-1}^k = \|(\widehat{\rho}_{x,n-1}^k, \widehat{\rho}_{y,n-1}^k)\|_2$, and $\rho_{x,n-1}^k$, $\rho_{y,n-1}^k$ are the Cartesian coordinates of the $k$-th track for emission number $n - 1$. Only tracks and blobs that meet constraint (24), are considered for the correlation

$$e_s^{k,i} = \left(\bar{\gamma}_n^i - H\widehat{\bar{\rho}}_{n|n-1}^k\right)^T \left(W_n\right)^{-1} \left(\underline{\gamma}_n^i - H\widehat{\underline{\rho}}_{n|n-1}^k\right). \quad (25)$$

The result is further thresholded,

$$e_s^{k,i} < G_s, \quad (26)$$

to ensure that a correct blob lies within the gate with a certain probability, which in turn is a function of the gate threshold $G_s$. Blobs that are not correlated with any track are initiated as a new track. Correlated blobs are further considered for a detection to track assignment.

### B. Detection to Track Assignment

For each chirp emission, the global assignment process determines the best association of blobs and tracks. Let $C \in \mathbb{R}^{N_t \times N_d}$ be the association matrix, where $N_t$ and $N_d$ are the number of tracks and blobs, respectively, at emission number $n$, respectively. $C(k, i)$ is given by

$$C(k, i) = \begin{cases} e_s^{k,i}, & \text{if track } k \text{ and blob } i \text{ fulfill (26)} \\ \infty, & \text{else} \end{cases}. \quad (27)$$

Let $\Upsilon$ be a set of tracks and blobs containing indices for possible assignments. The global assignment is obtained by finding the blobs-tracks assignment with the minimum global statistical distance, namely

$$\widehat{\Upsilon} = \arg\min_{(k,i) \in \Upsilon} \sum_k \sum_i C(k, i),$$
$$\text{s.t. } |\Upsilon| = \min(N_t, N_d). \quad (28)$$

To solve (28), we use the auction algorithm [28] due to its simplicity and low computational load. A blob that is not assigned to a track (in the case $N_d > N_t$) opens a new track. A track is removed if it has not been assigned to a blob at least $d_1$ times during the last $d_2$ emissions.

### C. Track Confirmation

A track is confirmed if the number of blobs assigned to it is greater than $N_c$. Formally,

$$\sum_i \mathbf{I}\left(e_s^{k,i} N_c\right), \quad (29)$$

where $\mathbf{I}$ is an indicator function that is 1 if the argument is true and 0 otherwise.

### D. Complexity Analysis

The complexity $C$ of our detection and tracking method per single emission is

$$C = O\left(n_x^{2.37} + UL_s|\Omega|\right). \quad (30)$$

This is composed of a complexity of $O(n_x^{2.37})$ for the debiased Kalman filter [29], where $n_x$ is the size of the state vector $\bar{\rho}$, and $O(UL_s|\omega|)$ stands for the 2D CFAR detection. While the complexity of the particle filter is $O(N_d^{n_x})$, and the complexity of the PHD filter is $O\big((n_x N_d)^3\big)$ [30]. In noisy environment, $N_d \gg n_x$, which implies that the proposed detection and tracking method has much lower complexity compared to the particle and PHD filters.

## VI. PERFORMANCE EVALUATION

In this section, we analyze the performance of our detection and tracking algorithm with both simulative and real data. We compare our method with benchmarks that focus on underwater acoustic detection and tracking. In particular, the methods in [31], referred below as *Lo*, and the scheme in [32], referred below as *Karoui*, which are commonly used underwater detection and tracking schemes designed specifically for noisy underwater environment.

The waveform $d(t)$ is a chirp signal with a bandwidth of 10 kHz and a duration of 0.01 s. The sampling frequency $f_s$ is set to 50 kHz. The merging thresholds $\psi_r$ and $\psi_\theta$ are set to 10 m and 6°, respectively. The standard deviations for the range and azimuth $\sigma_r$ and $\sigma_\theta$ are set to 0.3 m and 3°, respectively. The standard deviation of the process noise $\sigma_\zeta$ is set to $10^{-4}$ m/s². The gating thresholds for range and azimuth, $G_r$ and $G_\theta$, are set to 10 m and 10°, respectively, and the threshold value for gating, $G_s$, is set to 0.1. The threshold value for the track confirmation is set to $N_c = 5$. The track is deleted if the number of unassigned blobs exceeds 7 during 15 signal emissions.

### A. Results on Simulated Data

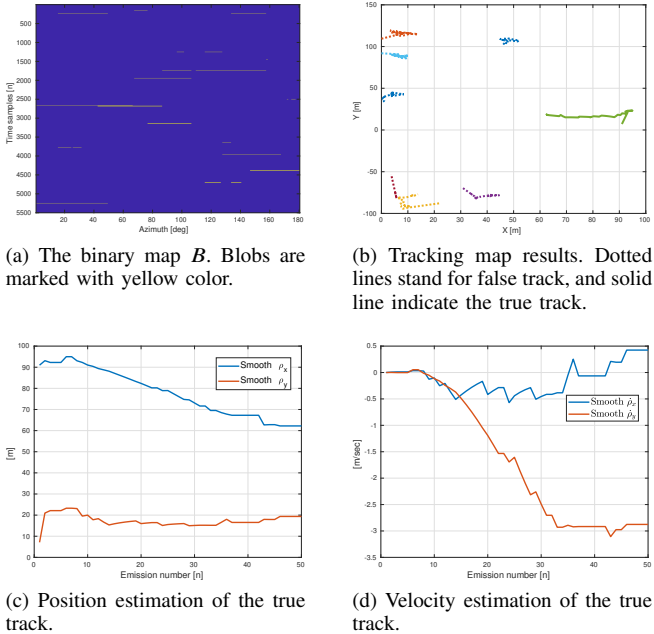

(a) The binary map $B$. Blobs are marked with yellow color.

(b) Tracking map results. Dotted lines stand for false track, and solid line indicate the true track.

(c) Position estimation of the true track.

(d) Velocity estimation of the true track.

Fig. 3. Example of detection and tracking results.

In this section, we analyze the performance of our proposed detection and tracking method using simulated data. We simulate a single target with an initial position of $\rho_x = 0$ m and $\rho_y = 100$ m and a velocity of $\dot\rho_x = 1$ m/s and $\dot\rho_y = -3$ m/s.

An example of the detection and tracking results can be found in Fig. 3. The binary map is shown in Fig. 3a, where the yellow regions indicate the detected blobs. Fig. 3b shows the track map. The solid line indicates the track that belongs to the real target and the dotted lines represent the false tracks. Fig. 3c and Fig. 3d show the position and velocity estimation of the target, respectively. We observe a convergence of velocity after 32 emissions. Fig. 4 shows an example of the samples received from the emission after beamforming. The target is marked with the arrow, and the SCR is 3 dB.

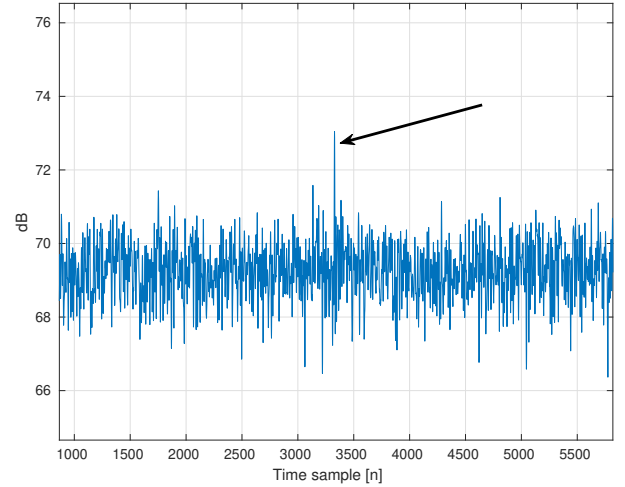

Fig. 4. An example of data samples after beamforming. Target is marked with the arrow.

*1) Comparison with the Benchmarks:* In this section, we compare the performance of our detection and tracking method with the Lo and Karoui benchmarks. For comparison, we measure the continuity of the estimated track. The track continuity is defined as the ratio between the duration of the track and the lifetime of the target. A track continuity equal to '1' would mean that the target is detected and continuously tracked for all signal emissions.

Results are shown for an SCR of 3 dB in Fig 5a and for an SCR of 5 dB in Fig 5b. For the latter figure, we note that the results for Lo are of less range than the other two schemes since it showed confirmed tracks only after a threshold corresponding to 4 false tracks. We observe that the results of Lo show the lowest track continuity. This is because this scheme does not merge closely spaced detection, which increases the false alarm rate and thus decreases the track continuity. We find that our method outperforms the benchmark for both tested SCR values and achieves a track continuity of more than 0.9 with less than two false traces. We notice the SCR has an impact on the tracking performance. For an SCR of 3 dB, our method achieves a track continuity of one with 7 false tracks, while, for SCR of 5 dB, the same tracking continuity is achieved with 2 false tracks.

*2) Confirmation of Threshold Analysis:* In this section, we analyze the continuity of tracks and the number of false tracks for different values of the track confirmation thresholds, $N_c$. In this analysis, we set the detection threshold $P_{fa}$ to 0.2 and the SCR to 3 dB. The results are shown in Fig. 6. When the confirmation threshold is small, more tracks are validated

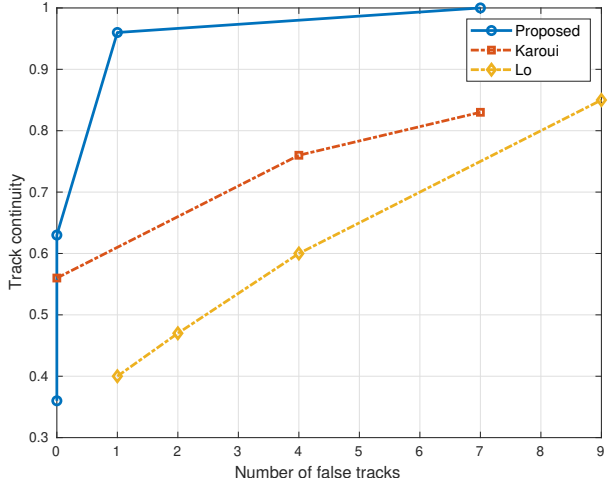

(a) Results are obtained with SCR = 3 dB.

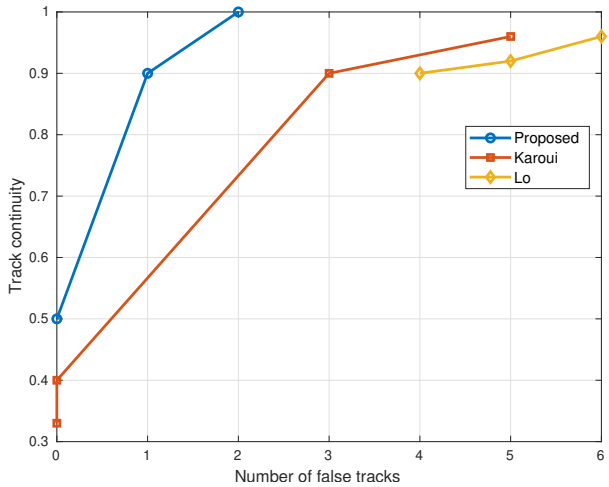

(b) Results are obtained with SCR = 5 dB.

Fig. 5. Track continuity vs. the number of false tracks to compare sensativity to the detection threshold. Results shown for our method and the two benchmark tracking schemes.

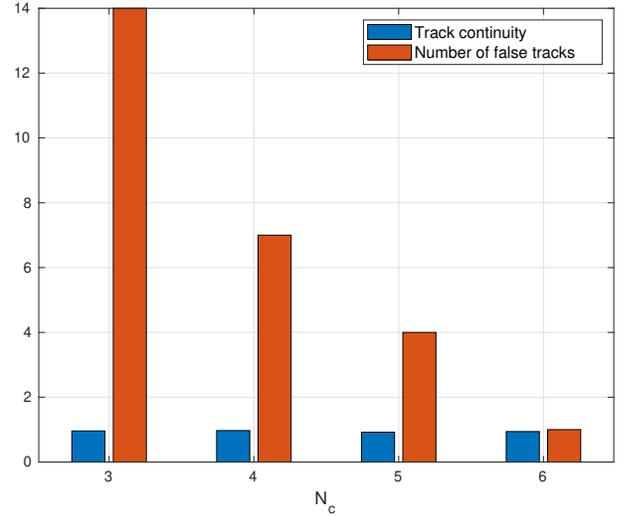

Fig. 6. Track continuity and number of false tracks obtained for different values of $N_c$. Results are obtained with SCR = 3 dB.

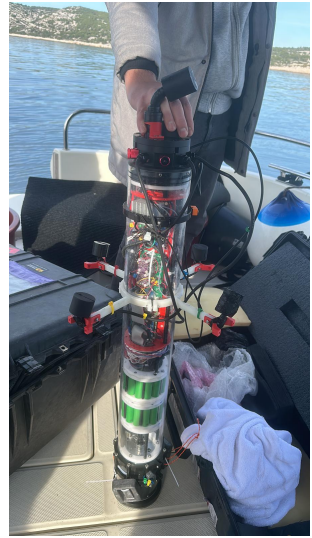

(a) The 4-element hydrophone array with the projector stationed above.

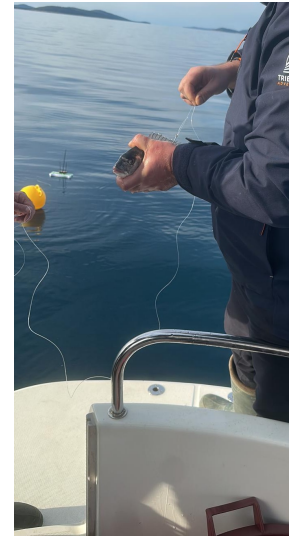

(b) A picture of one of the fish tested. The GPS and its attached surface float are shown in the water.

Fig. 7. Pictures from the sea experiment.

and the number of false tracks increases. However, the track associated with the true target is reported with less validation time. When the confirmation threshold is high, the validation time is also high and consequently the number of false tracks decreases, but the track associated with the true target is reported with a higher latency. As expected, increasing the confirmation threshold does not affect the continuity of the tracks, but rather reduces the number of false tracks.

### B. Sea Trial Results

*1) Experimental Setup:* To further validate the performance of our proposed detection and tracking method, we now present the results of a designated sea experiment conducted in Šibenik, Croatia, in January 2024. The experiment comprised a self-built floater with a self-built simultaneously sampled four hydrophone planar array and a self-built single projector (see Fig. 7a). The hydrophones were about 25 cm apart and the projector was positioned about 40 cm above the hydrophones. The measurements of all elements to a reference point on the floater were fed into the multidimensional scaling (MDS) algorithm [33] to obtain a 3D estimate of the position of the array. A schematic representation of our system can be found in Fig. 8a, where the hydrophones are labeled ch1 to ch4. The positions of ch1, ch2, ch3 and ch4 are (-38.5,0), (0,39.5), (40.5,0) and (0,-40.5) respectively. The effective beam pattern of this constellation is shown in Fig. 8b.

In our experiment, the floater was set to maintain a depth of 20 m with 1 m overshoot and the water depth was approximately 30 m. The position of the floater was monitored by attaching it to a small surface buoy with a GPS logger (2 m accuracy). The sea was calm, the bathymetry was approximately flat and the seabed consisted mainly of muddy sediment. The sound speed profile was 1500 m/sec in the first 3 m, due to a freshwater layer from the Krka River, and was

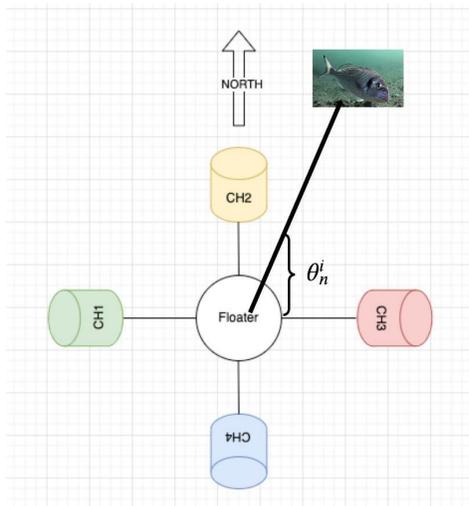

(a) A scheme of our platform with four hydrophones.

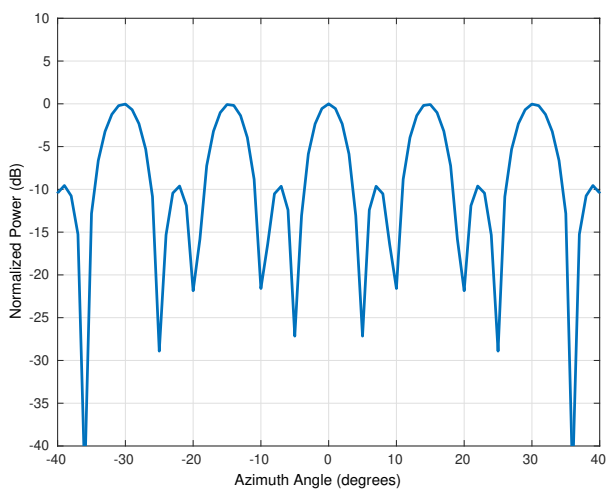

(b) Beam pattern in azimuth cut of the array.

Fig. 8. Scheme of the hydrophone array tested during the sea experiment.

fixed at 1520 m/sec below this layer. The emitted signals were linear frequency modulated chirps in the frequency band of 31 kHz-41 kHz and a duration of 10 ms. A guard interval of 1 second between adjacent emissions was used to suppress inter-pulse-interference, while the maximum distance for creating the distance-angle 2D matrices was 500 m. The sampling rate was set to 96 k samples per second at 2 bytes per sample.

We experimented with 2 gilt-head seabream as target fish. The fish were released at the same time near the floater buoy and swam freely away from it. The fish were attached with a 20 m fishing line to a float equipped with a GPS logger (2 m accuracy) to obtain information about the actual fish location. Due to the length of the fishing line and the inaccuracies in the GPS position, the inaccuracy in the true position of the fish is at most 24 m. We confirm that ethical approval for this experiment was obtained from the Ruđer Bošković Institute, Croatia, and the experiments were conducted in accordance with the approved guidelines and EU ethical regulations. The methods used followed the ARRIVE guidelines (https://arriveguidelines.org).

*2) Experiment Results:* Fig. 9 shows the results for the tracks obtained during the three experiments. The location estimates for fish #1 and for fish #2 are shown in Fig. 9a and Fig. 9c respectively. The velocity estimates for fish #1 and for fish #2 are shown in Fig. 9b and Fig. 9d respectively. Comparing the figures, we observe a different swimming behavior for the two fish, with fish #2 swimming faster and more smoothly. The two fish we experimented with for Exp 3 swam in different directions, but at approximately the same speed.

We note that the results of our tracking method are close to the GPS ground truth data, where in both cases the error was below the maximum ambiguity. For fish #1, convergence of position was achieved after 45 seconds, and the error in velocity estimation was about 0.2 m/s in both coordinates. Convergence of the position of fish #2 was achieved after 10 seconds and the velocity error was 0.2 m/s and 0.1 m/s for the X and Y coordinates, respectively.

Fig. 10 introduces a track continuity comparison for the sea trial results between our method and the two explored benchmarks. The track was associated with the true fish in the sea trial by the GPS data. Regarding the false detection, besides the tagged fish, more valid targets could exist. Thus, with no ground truth information for the number of false tracks, we measure the track continuity for different detection thresholds by changing the target false alarm rate as defined in (8). The results confirm our observations from the simulations in Fig. 5. Low track continuity is observed for the "Lo" benchmark, whereas our proposed approach show robuswtness to different choice of threshold.

## VII. Conclusion

We presented a detection and tracking method for underwater mobile targets. High acoustic reflections from clutter often increase the number of false alarms, which significantly affects the performance of underwater tracking. To overcome this challenge, we offered a track-before-detect mechanism that clusters sets of binary maps, and performs tracking on blobs rather that on individual detections. Towards an application of real-time analysis, we aimed for a low complexity tracking solution. Results on simulated data as well as on real data from a sea experiment reveal that our method effectively reduces the number of false tracks while keeping the track continuity high. Accurate tracking in location and velocity estimate is shown in the presence of low signal-to-clutter ratio that exists two benchmark schemes. Further work will extend this work for handling the spatial ambiguity of the array as well as to utilize the multiple arrivals reflected from the mobile target.

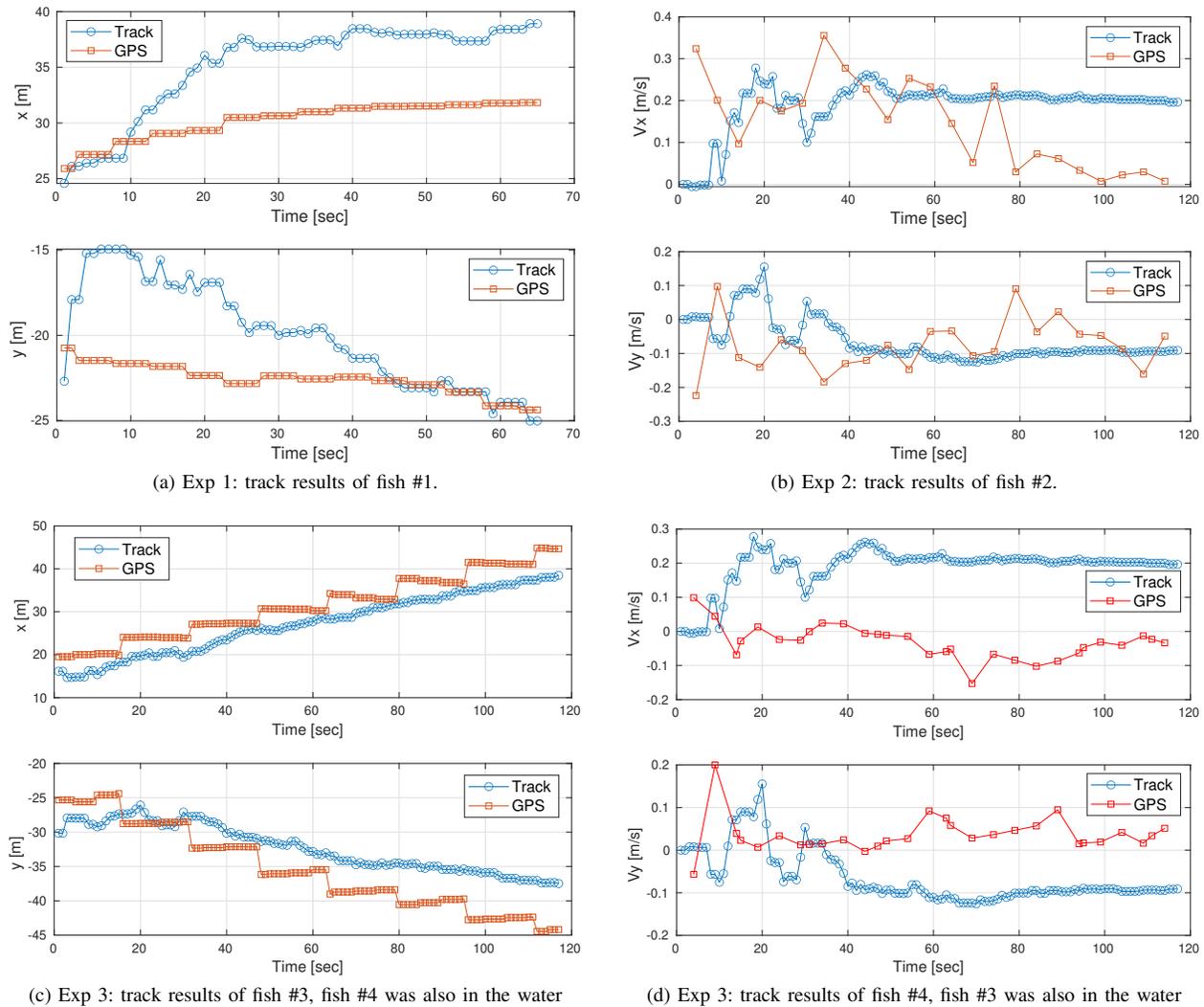

Fig. 9. Track results versus GPS ground truth data. Results are showing for the X-Y coordinates.

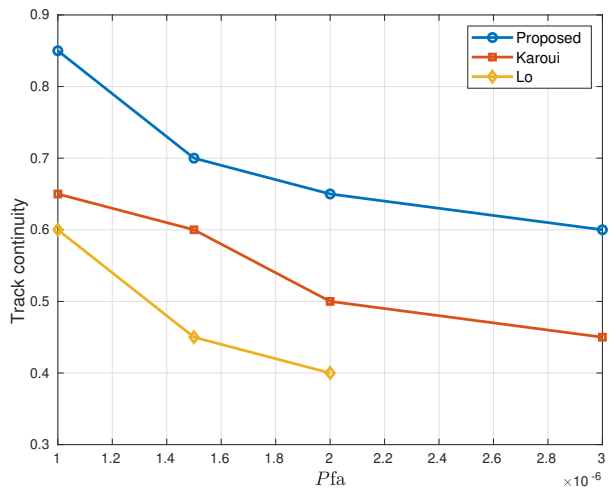

Fig. 10. Track continuity obtained for a detection threshold set by changing the target $P$fa values according to (8).